\documentclass{ametsocV6.1}

\usepackage{xcolor}
\definecolor{light-gray}{gray}{0.95}

\title{Lee-Wave Energy Sinks in Bottom-Intensified Flow: Reabsorption, Dissipation and Nonlinear Spectral Transfer}

\authors{
Yue Wu,\aff{a}\correspondingauthor{Yue Wu, ywu.ocean@gmail.com} 
Eric Kunze,\aff{b} 
Amit Tandon,\aff{c} 
Amala Mahadevan \aff{d} 
}

\affiliation{
\aff{a}{University of Michigan, Ann Arbor, Michigan}\\
\aff{b}{NorthWest Research Associates, Redmond, Washington}\\
\aff{c}{University of Massachusetts-Dartmouth, Dartmouth, Massachusetts}\\
\aff{d}{Woods Hole Oceanographic Institution, Woods Hole, Massachusetts}
}

\abstract{
Idealized numerical simulation is used to explore energy sinks for lee waves trapped in their bottom-intensified generating flow. In addition to the loss to explicit dissipation and reabsorption predicted by linear wave action conservation, indirect dissipation due to a nonlinear forward cascade by parametric subharmonic instability represents a significant sink that substantially reduces reabsorption. The partition of lee-wave energy loss between reabsorption and (explicit plus indirect) dissipation is independent of subgridscale damping parameterization. Remote dissipation of freely propagating internal waves generated by shear instability at the lee-wave critical layer proves to be small. A general parameterization for lee-wave dissipation of the balanced flow requires a more complete exploration of the parameter space.
}

\begin{document}
\maketitle


\section{Introduction}

Ocean lee waves are internal gravity waves (IGWs) generated by stably stratified flow over bathymetry \citep{Bell1975}. 
Their generation exerts wave drag and extracts energy from the balanced flow. Their propagation transports this energy through wave-fluxes. When they break, this energy cascades to turbulent dissipation and mixing, contributing to maintenance of ocean stratification, the meridional overturning circulation and dissipation of the largescale circulation (e.g., \citealt{Melet2014,MacKinnon2017,Kunze2017}). 

Lee waves generated by a steady flow have Eulerian frequencies $\omega_E=0$ in a fixed reference frame to maintain stationary phase with respect to topography. 
Their intrinsic or Lagrangian frequencies $\omega_I=|kU_0|$, where $k$ is the alongstream topographic wavenumber and $|U_0|$ the near-bottom flow speed. 
For typical ocean abyssal flow speeds $|U_0|=O$(0.1) m s$^{-1}$, topography with wavelengths of $O$(1--10) km generates lee waves with $|f|<|kU_0|<N$ where $f$ is the Coriolis frequency and $N$ buoyancy frequency; these wavelengths are not resolved by satellite bathymetry \citep{Kunze2004}. Outside this band, the response is evanescent. 

Lee-wave generation can be expressed as $\rho_0U_0h^2(k^2U_0^2-f^2)^{1/2}(N^2-k^2U_0^2)^{1/2}$, where $\rho_0$ is the reference density and $h$ topographic height. Bottom flow $U_0(x,y)$ can be modulated by low topographic wavenumbers associated with critical near-inertial and subinertial topography ($|kU_0|<|f|$ with $1/k>10$ km) whose topographic steepness $\epsilon=mh=khN/|k^2U_0^2 - f^2|^{1/2} > 1$ with $m$ being the lee-wave vertical wavenumber. A complication is that ocean general circulation models (OGCMs) tend to underestimate abyssal currents \citep{Scott2010}, possibly because they do not resolve these near-inertial and subinertial topographic interactions that modify the near-bottom flow over height scales $1/m$ \citep{Hogg1973, Klymak2010}.

In the water column, lee waves cannot escape their generating current but are trapped inside the $|U|=|f/k|$ isotach, reflecting laterally from turning points on the cross-stream boundaries of the flow where the cross-stream wavenumber passes through zero, and stalling at vertical critical layers where their vertical wavelengths and group velocities shrink and velocities amplify. Critical-layer stalling acts to increase the gradient Froude number $Fr_g = |\mathbf{v}_z|/N$ (where $\mathbf{v}_z$ is vertical shear) until instability ensues, leading to turbulent production \citep{Kunze1985a, Kunze1995} and possibly free-wave radiation (e.g., \citealt{Pham2009,Zemskova2021}).

Recently developed OGCM parameterizations for lee-wave generation by balanced circulation (e.g., \citealt{Nikurashin2011, Melet2014, Melet2015}) have assumed that all lee-wave generation is lost to local turbulent dissipation and mixing. 
However, microstructure measurements of turbulent dissipation at two major lee-wave generation sites in the Southern Ocean fall short of linear lee-wave generation predictions by as much as an order of magnitude, in particular, in bottom-intensified flows \citep{Sheen2013, Waterman2013, Waterman2014, Cusack2017, Cusack2020}.
Several mechanisms for this so-called ``suppression of turbulence" are summarized in \cite{Waterman2014} and \cite{Kunze2019}. Here, we explore (i) reabsorption of lee-wave energy back into bottom-intensified flow \citep{Kunze2019, Wu2023}, and (ii) generation of freely-propagating IGWs (free waves; $\omega_E \neq 0$) that can escape a localized generating current to dissipate remotely \citep{Kunze1995, Wright2014}.

Reabsorption of lee-wave energy in bottom-intensified flow has been studied theoretically by \cite{Kunze2019} and numerically by \cite{Wu2023}. Based on wave action conservation, \cite{Kunze2019} partitioned lee-wave energy into dissipative and reabsorptive fractions, where the minimum dissipative fraction is $|f/kU_0|$ for waves that reach the lee-wave critical-layer and the remaining fraction is reabsorbed by the bottom-intensified flow as the lee waves propagate upward into the water column and weaken background flow. The theory does not consider nonlinear lee waves which may break before $|kU_0| \downarrow |f|$, enhancing the dissipative fraction at the expense of the reabsorptive fraction. The numerical simulations of \cite{Wu2023} found that the reabsorptive fraction fell short though the explicit dissipation appeared to be consistent with wave action conservation predictions.
While the idealized study of \cite{Wu2023} quantified the fractions of lee-wave generation lost to dissipation, reabsorption and nonlinear transfer in a bottom-intensified jet, it had limitations. First, it employed a strong vertical viscosity and diffusivity of $O(10^{-3}$ m$^2$ s$^{-1})$ to damp lee-wave instability near the critical layer to ensure a steady state. This eliminated temporal instabilities and generation of free waves so that mechanism (ii) could not be studied. Second, dissipation in numerical models relies on parameterization of turbulent viscosities, but the theoretical dissipative fraction $|f/kU_0|$ only depends on the lee-wave intrinsic frequency at generation and at breaking. It is not clear whether the theoretical dissipative fraction still holds if a different turbulent parameterization is employed. Third, a nonlinear transfer sink was substantial though not part of linear wave action conservation theory, and hence not interpreted correctly.

In this study, we extend the recent numerical modeling of lee-wave generation, propagation, interaction, reabsorption and dissipation in a bottom-intensified jet \citep{Wu2023} by reducing vertical gridscale damping by an order of magnitude to allow shear instabilities and generation of free waves near the critical layer, where the gradient Froude number exceeds its critical value of 2.
The goal is to (1) quantify the relative roles of remote dissipation by free waves versus local dissipation by trapped lee waves, (2) explore wave-mean and wave-wave interactions, and (3) test the sensitivity of lee-wave energy sink partition to the subgridscale damping parameterization.
The numerical results demonstrate that reabsorption of lee-wave energy in bottom-intensified flow has an $O(1)$ effect so may be key to the measured turbulence shortfall, while remote dissipation in the form of free waves is negligible for the tested parameter range.
While lee waves explicitly dissipate less with reduced viscosity, they transfer a significant fraction of their energy to free waves.
Self-advection of both lee and free waves acts as an additional nonlinear sink which is interpreted as a cascade to unresolved small scales and indirect dissipation. For both the steady \citep{Wu2023} and free-wave-perturbed (this paper) cases, roughly 50\% of the energy is reabsorbed and 50\% is lost to explicit dissipation plus the nonlinear sink. The total (explicit and indirect nonlinear) dissipative fraction is considerably higher than $|f/kU_0| = 15\%$ predicted by linear wave action conservation \citep{Kunze2019}, implying that nonlinear wave-wave interactions reduce the reabsorptive fraction in favor of dissipation.

The remainder of the paper is organized as follows. Section 2 describes the model setup. Section 3 introduces a triple energy conservation decomposition for the jet, lee waves and free waves. Section 4 presents the simulation results, and Section 5 discusses energy budgets. Section 6 is the summary and discussion.

\section{Model Setup}

Following \cite{Wu2023}, the regional Process Study Ocean Model (PSOM, \citealt{Mahadevan1996a, Mahadevan1996b}) is configured with a bottom-intensified, laterally confined jet over sinusoidal topography. The zonally periodic domain is 5 km in the alongstream (zonal) direction, 20 km in the across-stream (meridional) direction and 2 km in depth with 0.1-km grid spacing in the horizontal and 4-m in the vertical. 
The south, north and bottom boundaries are rigid and free-slip, while the top boundary is a free surface.
The Coriolis frequency $f=-1.3\times10^{-4}$ rad s$^{-1}$ which corresponds to 63.4°S latitude, and buoyancy frequency $N=10^{-3}$ rad s$^{-1}$ is representative of the abyssal Southern Ocean. 

The background state consists of a stable zonal jet in thermal-wind balance. The jet is bottom-intensified and meridionally confined (Figure~\ref{fig_u}a), with maximum speed $|U_0|=0.18$ m s${^{-1}}$ decaying over 1900 m vertically and 7 km meridionally so that both its Rossby number and gradient Froude number are close to 0.1. 
Bottom intensification mimics conditions for which \cite{Waterman2014} reported that turbulent dissipation fell short of linear lee-wave generation predictions. Bottom intensification on vertical scales $O(U/N)$ is expected for flow over subinertial $(|kU|<|f|)$ topography \citep{Hogg1973, Klymak2010}. The jet is isolated from the meridional boundaries so lee waves do not interact with them. 

The topography is monochromatic with alongstream wavenumber $k_0=5.23 \times 10^{-3}$ rad m$^{-1}$ (wavelength $\lambda_x=1.2$ km and lee-wave intrinsic frequency $|k_0U_0| \sim 0.9N$) and amplitude $a = 5$ m so that the topographic Froude number $Fr_t = aN/U_0 = 0.05$ is subcritical, implying linear generation of lee waves \citep{Bell1975,Nikurashin2010a}. 
The critical-layer depth where $|k_0U|\approx |f|$ is 800 m along the central axis of the jet (Figure~\ref{fig_u}). The jet is maintained in steady state against damping by lee-wave generation and dissipation using a restoring force \citep{Wu2023} to mimic replenishment of the jet by large-scale forcing. Unstable shear, wave-breaking, turbulence, as well as possible free-wave radiation, are expected near the critical depth \citep{Jones1967,Kunze1985a, Kunze1995}. IGW radiation from unstable shear layers has been observed in the atmosphere (e.g., \citealt{Einaudi1978, Holton1995, Rosenlof1996}), in the ocean (e.g, \citealt{Moum1992, Sun1998}), numerically (e.g., \citealt{Sutherland1994, Skyllingstad1994, Sutherland1996, Smyth2002, Tse2003, Sutherland2006, Basak2006, Pham2009, Nikurashin2010a, Zemskova2021}), and in the lab (e.g., \citealt{Strang2001}). 

While wave-breaking and turbulence cannot be resolved in a regional numerical model so that their effects must be parameterized, here vertical viscosities and diffusivities are reduced to 10$^{-4}$ m$^2$s$^{-1}$ with a corresponding grid damping timescale of 0.5 days; these viscosities and diffusivities are a factor of 40 smaller than those used by Wu et al. (2023) to allow shear instability and IGW radiation. Biharmonic horizontal damping $A_h=K_h = 125$ m$^4$s$^{-1}$ (for momentum and density) with a grid damping timescale of 0.6 days is used as in \cite{Wu2023}. The advantage of using biharmonic instead of Laplacian damping in the horizontal is that hyper-viscosity/diffusivity can more efficiently eliminate spurious gridscale variance near meridional boundaries while preserving scales of interest.

\section{Triple decomposition and energetics}

The Eulerian governing equations for a Boussinesq fluid on an $f$-plane are 
\begin{equation}
\frac{D u_i}{Dt} -\varepsilon_{ij}fu_j + \rho_0^{-1}\frac{\partial p}{\partial x_i} - b \delta_{i3} = \mathcal{F}^m +\text{viscous terms}, \label{T1}
\end{equation}
\begin{equation}
\frac{Db}{Dt} +N^2 u_3 = \mathcal{F}^b + \text{diffusive terms}, \label{T2}
\end{equation}
\begin{equation}
\frac{\partial u_i}{\partial x_i} = 0,
\end{equation}
where $D/Dt=\partial/\partial t+ u_j (\partial/\partial x_j)$ is the material or Lagrangian time derivative, $i$ and $j$ run from 1 to 3, $\varepsilon_{ij}=1\text{ }(-1)$ if $(i,j)$ is an even (odd) permutation of $(1,2)$ and zero otherwise, and $\delta_{ij}=1$ if $i=j$ and zero otherwise. 
$u_i=(u_1,u_2,u_3)=(u,v,w)$ is the three-dimensional (3D) velocity, and $b=-g[\rho-\rho^*(z)]/\rho_0$ buoyancy anomaly in which $g$ is gravitational acceleration, $\rho$ density, $\rho_0=1027$ kg~m$^{-3}$ reference density, and $\rho^*(z)$ rearranged density with flattened isopycnals to achieve a state of a minimum potential energy \citep{Winters1995,Klymak2018}. 
$p$ is the pressure deviation with respect to the minimum PE state with $\rho^*(z)$, and $N=\sqrt{-g\frac{d\rho^*(z)}{dz}/\rho_0}$ the buoyancy frequency. 
$\mathcal{F}^m$ and $\mathcal{F}^b$ are external forcing for momentum and buoyancy, respectively. 

To quantify wave-mean and wave-wave interactions in a steady state without the decay of the jet due to loss to lee-wave generation and dissipation, the simulation is forced to maintain a quasi-steady state so that both the jet and lee waves are steady while time-dependent perturbations associated with free waves can propagate. 
$\mathcal{F}^m$ and $\mathcal{F}^b$ take the form $D\Psi/Dt = -(1/\tau)\left(\bar{\Psi} - \bar{\Psi}_\text{ini}\right)$, where $\bar{\Psi}$ represents the zonally averaged instantaneous velocity or density field, and $\bar{\Psi}_\text{ini}$ the zonally averaged initial field. 
Results are not sensitive to the restoration timescale $\tau$, which is 1 day. The restoration will also act on any $k = 0$ motions so will damp inertial free waves though this sink proves to be small in the result section.

Multiplying \eqref{T1} by $u_i$ and \eqref{T2} by $b/N^2$, then summing gives the \emph{total energy conservation}
\begin{equation}
\frac{\partial}{\partial t}\left(\text{KE+APE}\right) = \underbrace{ -u_j\frac{\partial}{\partial x_j}\left(\text{KE+APE}\right) }_\text{total advection} - \underbrace{\rho_0^{-1}\frac{\partial}{\partial x_i}\left(u_i p\right) }_\text{total pressure work} + \text{restoration} + \text{dissipation}, \label{T3}
\end{equation}
where kinetic energy $\text{KE} = \frac{1}{2}u_i^2 =\frac{1}{2}\left(u^2+v^2+w^2\right)$ and available potential energy $\text{APE} = \frac{1}{2} N^{-2} b^2$. With $\rho^*$ subtracted from the numerator of buoyancy anomaly $b$, this definition of APE has contributions from the jet, lee waves and free waves. 
Restoration is equivalent to continuously replenishing KE and APE in the system so that the left-hand side (LHS) of \eqref{T3} is zero.

\subsection{Triple decomposition}

To separate the jet $(\bar{~})$, lee waves $(\tilde{~})$ and free waves $(')$, a triple decomposition is applied. We use angle brackets and overbars to denote time and zonal averages, respectively. 
We first take a time-average of the total velocity and buoyancy to separate the steady fields (i.e., the jet plus lee waves) from time-dependent perturbations (i.e., free waves $\mathbf{u}'$ and $b'$ whose time-averages $\langle \mathbf{u}'\rangle = \langle b'\rangle = 0$). We then take a zonal average of the steady fields to separate the jet ($\bar{\mathbf{u}}$ and $\bar{b}$, invariant in both time and zonal direction) and lee waves ($\tilde{\mathbf{u}}$ and $\tilde{b}$, steady but zonally varying with topography whose zonal averages $\overline{\tilde{\mathbf{u}}} = \overline{\tilde{b}} = 0$)
\[\mathbf{u} = \overbrace{\underbrace{\bar{\mathbf{u}}}_\text{mean}+\underbrace{\tilde{\mathbf{u}}}_\text{lee waves}}^{= \text{ steady }\langle \mathbf{u} \rangle }
+\underbrace{\mathbf{u}'}_\text{free waves},\text{    }
b = \overbrace{\underbrace{\bar{b}}_\text{mean}+\underbrace{\tilde{b}}_\text{lee waves}}^{= \text{ steady }\langle b \rangle }
+\underbrace{b'}_\text{free waves}.
\]
Bold symbols denote vectors.

\subsection{Mean energy conservation}

Decomposition and linearization of the nonlinear advective terms in \eqref{T1} and \eqref{T2} give rise to additional terms associated with nonzero wave fluxes $\overline{\tilde{u}_i\tilde{u}_j}$, $\overline{\tilde{b} \tilde{u}_j}$, $\overline{\langle u'_i u'_j\rangle}$ and $\overline{\langle b' u'_j\rangle}$. The mean momentum and buoyancy equations are \citep{Reynolds1972}
\begin{align}
\frac{\partial \bar{u}_i}{\partial t}+\bar{u}_j\frac{\partial \bar{u}_i}{\partial x_j} -\varepsilon_{ij}f\bar{u}_j + \rho_0^{-1}\frac{\partial \bar{p}}{\partial x_i} -\bar{b}\delta_{i3}
=-\frac{\partial}{\partial x_j} \left(\overline{\tilde{u}_i \tilde{u}_j}+ \overline{\langle u'_i u'_j\rangle} \right) + \bar{\mathcal{F}}^m + \text{mean viscous terms}, \label{M1} \\
\frac{\partial \bar{b}}{\partial t}+\bar{u}_j\frac{\partial \bar{b}}{\partial x_j} + N^2\bar{u}_3
=-\frac{\partial}{\partial x_j} \left(\overline{\tilde{b} \tilde{u}_j}+ \overline{\langle b' u'_j\rangle} \right) +\bar{\mathcal{F}}^b + \text{mean diffusive terms}, \label{M2}
\end{align}
\[\bar{\mathcal{F}}^m=-\frac{1}{\tau}(\bar{u}_1-\bar{u}_{1,\text{ini}}), \text{    }\bar{\mathcal{F}}^b=-\frac{1}{\tau}(\bar{b}-\bar{b}_{\text{ini}}),
\]
\citep{Reynolds1972} where $\bar{u}_{1,\text{ini}}$ and $\bar{b}_{\text{ini}}$ are the initial zonal velocity and buoyancy fields of the jet, respectively. 

Taking the product of \eqref{M1} with $\bar{u}_i$, multiplying \eqref{M2} by $N^{-2}\bar{b}$, and summing gives the \emph{mean energy conservation}
\begin{align} \label{M3}
\frac{\partial}{\partial t}\left(\text{MKE}+\text{MAPE}\right) = \underbrace{\left[\frac{\partial \bar{u}_i}{\partial x_j}\left(\overline{\tilde{u}_i \tilde{u}_j}+ \overline{\langle u'_i u'_j\rangle} \right) +N^{-2}\frac{\partial \bar{b}}{\partial x_j}\left(\overline{\tilde{b} \tilde{u}_j}+\overline{\langle b' u'_j\rangle} \right)\right]}_\text{exchange with lee- and free waves} \nonumber\\
\underbrace{-\frac{\partial}{\partial x_j} \left[ \bar{u}_i \left(\overline{\tilde{u}_i \tilde{u}_j}+ \overline{\langle u'_i u'_j\rangle} \right)+N^{-2} \bar{b} \left(\overline{\tilde{b} \tilde{u}_j}+ \overline{\langle b' u'_j\rangle} \right)\right]}_\text{lee- and free-wave drag} 
\underbrace{-\bar{u}_j\frac{\partial}{\partial x_j}\left(\text{MKE}+\text{MAPE}\right)}_\text{mean self-advection}\nonumber\\
\underbrace{-\rho_0^{-1}\frac{\partial}{\partial x_i}\left(\bar{u}_i\bar{p}\right)}_\text{mean pressure-work}
\underbrace{-\frac{1}{\tau} \left[\bar{u}_1 (\bar{u}_1-\bar{u}_{1,\text{ini}}) + N^{-2}\bar{b}(\bar{b}-\bar{b}_\text{ini}) \right]}_\text{restoration of mean} +\text{mean dissipation}, 
\end{align}
where mean kinetic energy MKE $ = \frac{1}{2}\bar{u_i}^2 = \frac{1}{2} \left(\bar{u}^2+\bar{v}^2+\bar{w}^2\right)$ and mean available potential energy MAPE $=\frac{1}{2}{N^{-2}} \bar{b}^2$. The first group of terms on the right-hand side (RHS) of \eqref{M3} is mean energy exchange with lee waves and free waves.
The second group of terms is transmission of lee-wave drag from the bottom boundary into the interior and free-wave drag.
The third term is advection of MKE and MAPE by mean velocities, the fourth term mean pressure-work, the fifth group of terms restoration of the mean, and the last term dissipation of the mean. 

\subsection{Lee-wave energy conservation}
The lee-wave momentum and buoyancy equations are
\begin{align}
\frac{\partial \tilde{u}_i}{\partial t} + \bar{u}_j\frac{\partial \tilde{u}_i}{\partial x_j} -\varepsilon_{ij}f\tilde{u}_j +\rho_0^{-1}\frac{\partial \tilde{p}}{\partial x_i} -\tilde{b}\delta_{i3}
= -\tilde{u}_j\frac{\partial \bar{u}_i}{\partial x_j}+\frac{\partial}{\partial x_j} \left(\overline{\tilde{u}_i \tilde{u}_j}-\tilde{u}_i \tilde{u}_j + \overline{\langle u'_i u'_j\rangle}-\langle u'_i u'_j\rangle \right) \nonumber\\
+\tilde{\mathcal{F}}^m + \text{lee-wave viscous terms}, \label{L1}\\
\frac{\partial \tilde{b}}{\partial t} + \bar{u}_j\frac{\partial \tilde{b}}{\partial x_j} + N^2\tilde{u}_3
= -\tilde{u}_j\frac{\partial \bar{b}}{\partial x_j}+\frac{\partial}{\partial x_j} \left(\overline{\tilde{b} \tilde{u}_j}-\tilde{b} \tilde{u}_j + \overline{\langle b' u'_j\rangle}-\langle b' u'_j\rangle \right) \nonumber\\
+\tilde{\mathcal{F}}^b + \text{lee-wave diffusive terms}. \label{L2}
\end{align}
where $\tilde{\mathcal{F}}^m=\tilde{\mathcal{F}}^b=0$ because restoration only acts on zonally-averaged fields. 

To obtain the equation for lee-wave energy conservation, \eqref{L1}--\eqref{L2} are multiplied by $\tilde{u}_i$ and $N^{-2}\tilde{b}$, respectively, zonally averaged and summed
\begin{align} \label{L3}
\frac{\partial}{\partial t}\left(\overline{\text{LKE}+\text{LAPE}}\right)= 
\underbrace{\left( -\frac{\partial \bar{u}_i}{\partial x_j} \overline{\tilde{u}_i \tilde{u}_j} + \overline{\frac{\partial \tilde{u}_i}{\partial x_j}\langle u'_i u'_j\rangle}-N^{-2}\frac{\partial \bar{b}}{\partial x_j} \overline{\tilde{b} \tilde{u}_j} + N^{-2}\overline{\frac{\partial \tilde{b}}{\partial x_j}\langle b' u'_j\rangle} \right)}_\text{exchange with mean and free waves} \nonumber\\
\underbrace{-\frac{\partial}{\partial x_j} \left(\overline{\tilde{u}_i\langle u'_i u'_j\rangle}+ N^{-2}\overline{\tilde{b}\langle b' u'_j\rangle}\right)}_\text{free-wave drag} 
\underbrace{-\bar{u}_j\frac{\partial}{\partial x_j}\left(\overline{\text{LKE}+\text{LAPE}}\right)}_\text{advection by mean}
\underbrace{-\frac{\partial}{\partial x_j} \left(\frac{1}{2} \overline{ \tilde{u}_i \tilde{u}_i \tilde{u}_j}+\frac{1}{2}N^{-2}\overline{ \tilde{b} \tilde{b} \tilde{u}_j}\right)}_\text{lee-wave self-advection} \nonumber\\
\underbrace{-\rho_0^{-1}\frac{\partial}{\partial x_i} \left(\overline{\tilde{u}_i\tilde{p}}\right)}_\text{lee-wave pressure-work}+\text{lee-wave dissipation}, 
\end{align}
where lee-wave kinetic energy LKE $ = \frac{1}{2}\tilde{u_i}^2 = \frac{1}{2} \left(\tilde{u}^2+\tilde{v}^2+\tilde{w}^2\right)$ and lee-wave available potential energy LAPE $=\frac{1}{2}{N^{-2}} \tilde{b}^2$. Lee-wave energy is advected by the mean flow and lee waves themselves (lee-wave self-advection), among which the latter is higher-order but not necessarily small for nonlinear waves. 
Lee-wave pressure-work at the bottom boundary is often referred to as lee-wave generation, or energy conversion from balanced flow into lee waves. 

\subsection{Free-wave energy conservation}

Free-wave momentum and buoyancy equations are obtained by subtracting the mean and lee-wave equations \eqref{M1}, \eqref{M2}, \eqref{L1} and \eqref{L2} from the total equations
\begin{align}
\frac{\partial u'_i}{\partial t} + \bar{u}_j\frac{\partial u'_i}{\partial x_j} -\varepsilon_{ij}f u'_j +\rho_0^{-1}\frac{\partial p'}{\partial x_i} - b'\delta_{i3}
= -\left(\tilde{u}_j\frac{\partial u'_i}{\partial x_j}+u'_j\frac{\partial \bar{u}_i}{\partial x_j}+u'_j\frac{\partial \tilde{u}_i}{\partial x_j} \right)+ \frac{\partial}{\partial x_j} \left(\langle u'_i u'_j\rangle - u'_i u'_j\right) \nonumber\\
+\mathcal{F}'^m + \text{free-wave viscous terms}, \label{F1}\\
\frac{\partial b'}{\partial t} + \bar{u}_j\frac{\partial b'}{\partial x_j} +N^2 u'_3
= -\left(\tilde{u}_j\frac{\partial b'}{\partial x_j}+u'_j\frac{\partial \bar{b}}{\partial x_j}+u'_j\frac{\partial \tilde{b}}{\partial x_j} \right)+ \frac{\partial}{\partial x_j} \left(\langle b' u'_j\rangle - b' u'_j\right) \nonumber\\
+\mathcal{F}'^b +\text{free-wave diffusive terms}, \label{F2}
\end{align}
\[\mathcal{F}'^m=-\frac{1}{\tau}\bar{u}'_1,\text{    } \mathcal{F}'^b=-\frac{1}{\tau} \bar{b}'.\]
where $\mathcal{F}'^m$ and $\mathcal{F}'^b$ are the restoration of the zonal velocity and buoyancy perturbations and are not necessarily zero due to the $k=0$ free waves.

Free-wave energy conservation is derived by multiplying \eqref{F1}--\eqref{F2} by $u'_i$ and $N^{-2} b'$, respectively, time and zonal averaging, and summing\begin{align} \label{F3}
\frac{\partial}{\partial t}\left(\overline{\langle\text{FKE}+\text{FAPE}\rangle} \right)= \underbrace{\left( -\frac{\partial \bar{u}_i}{\partial x_j} \overline{\langle u'_i u'_j\rangle} - \overline{\frac{\partial \tilde{u}_i}{\partial x_j}\langle u'_i u'_j\rangle} -N^{-2}\frac{\partial \bar{b}}{\partial x_j} \overline{\langle b' u'_j\rangle} - N^{-2}\overline{\frac{\partial \tilde{b}}{\partial x_j}\langle b' u'_j\rangle} \right) }_\text{exchange with mean and lee waves} \nonumber\\
\underbrace{-\bar{u}_j\frac{\partial}{\partial x_j}\left(\overline{\langle\text{FKE}+\text{FAPE}\rangle} \right)}_\text{advection by mean}
\underbrace{-\overline{\tilde{u}_j \frac{\partial}{\partial x_j} \left(\langle\text{FKE}+\text{FAPE}\rangle\right) }}_\text{advection by lee waves}
\underbrace{-\frac{\partial}{\partial x_j} \left(\frac{1}{2}\overline{\langle u'_i u'_i u'_j\rangle}+\frac{1}{2}N^{-2}\overline{\langle b' b' u'_j\rangle} \right)}_\text{free-wave self-advection} \nonumber\\ 
\underbrace{-\rho_0^{-1}\frac{\partial}{\partial x_i} \left(\overline{\langle u'_ip'\rangle}\right)}_\text{free-wave pressure-work}
\underbrace{-\frac{1}{\tau} \left[\langle \bar{u}'_1\bar{u}'_1\rangle + N^{-2}\langle \bar{b}'\bar{b}'\rangle \right]}_\text{restoration of free waves} +
\text{free-wave dissipation},
\end{align}
where free-wave kinetic energy FKE $=\frac{1}{2} u'^2_i = \frac{1}{2}\left(u'^2+v'^2+w'^2\right)$ and free-wave available potential energy FAPE $=\frac{1}{2}N^{-2} b'^2$. 
Free-wave energy is advected by the velocity of the mean, lee waves, and free waves themselves (free-wave self-advection), amongst which the last is higher-order and associated with nonlinearity of free waves. 
Restoration acts as an artificial damping on $k=0$ free waves that are near-inertial with little buoyancy.
 
\subsection{Summary of wave-mean and wave-wave interactions}

Energy conservation equations \eqref{M3}, \eqref{L3} and \eqref{F3} summarize wave-mean and wave-wave interactions in our simulation.
For each of the three fields -- mean jet, lee waves and free waves -- the time rate of change of kinetic plus available potential energy is governed by six forcings, i.e., exchange, drag, advection, pressure-work, restoration and dissipation.

The first three forcings (i.e., exchange, drag and advection) are interactive, connecting the three fields. They are derived by decomposition of the nonlinear advection term in the total equation. 
Exchange always appears in pairs, summing to zero as one field loses energy and the other gains the same amount to produce no net energy.
Drag acts on the lower-frequency field due to fluxes of the higher-frequency fields (the mean is subject to drag from both lee and free waves, lee waves are subject to drag from free waves, and no drag acts on free waves). 
Advection is complicated with energy in one field advected by the velocity of lower-frequency fields as well as the field itself (i.e., self-advection). Advection by lower-frequency fields is easy to interpret, representing transport of energy by background velocity. Self-advection takes the form of triple co-variance and is a higher-order term in \eqref{M3}, \eqref{L3}, and \eqref{F3}. It can be related to spectral energy transfer in wavenumber space due to nonlinear wave-wave interactions \citep{Skitka2023, Wu2023b} that occurs at a longer timescale than the linear timescale (i.e., the wave period) in weakly turbulent flow \citep{Nazarenko2011, Lvov2012}. 
Here, it is necessary to distinguish leading- and higher-order nonlinear interactions: exchange, drag and advection by lower-frequency fields are leading-order terms in \eqref{L3} and \eqref{F3} that connect different fields and represent energy transfer from one field to another, while lee- and free-wave self-advection are higher-order terms that represent energy transport in physical and spectral space. 

The remaining three forcings (i.e., pressure-work, restoration and dissipation) are noninteractive but due to boundary effects (i.e., bottom generation), external forcing (i.e., restoration) and damping. 

\section{Simulation results}

\begin{figure*}[hbp]
\centering
 \noindent\includegraphics[width=\textwidth,angle=0]{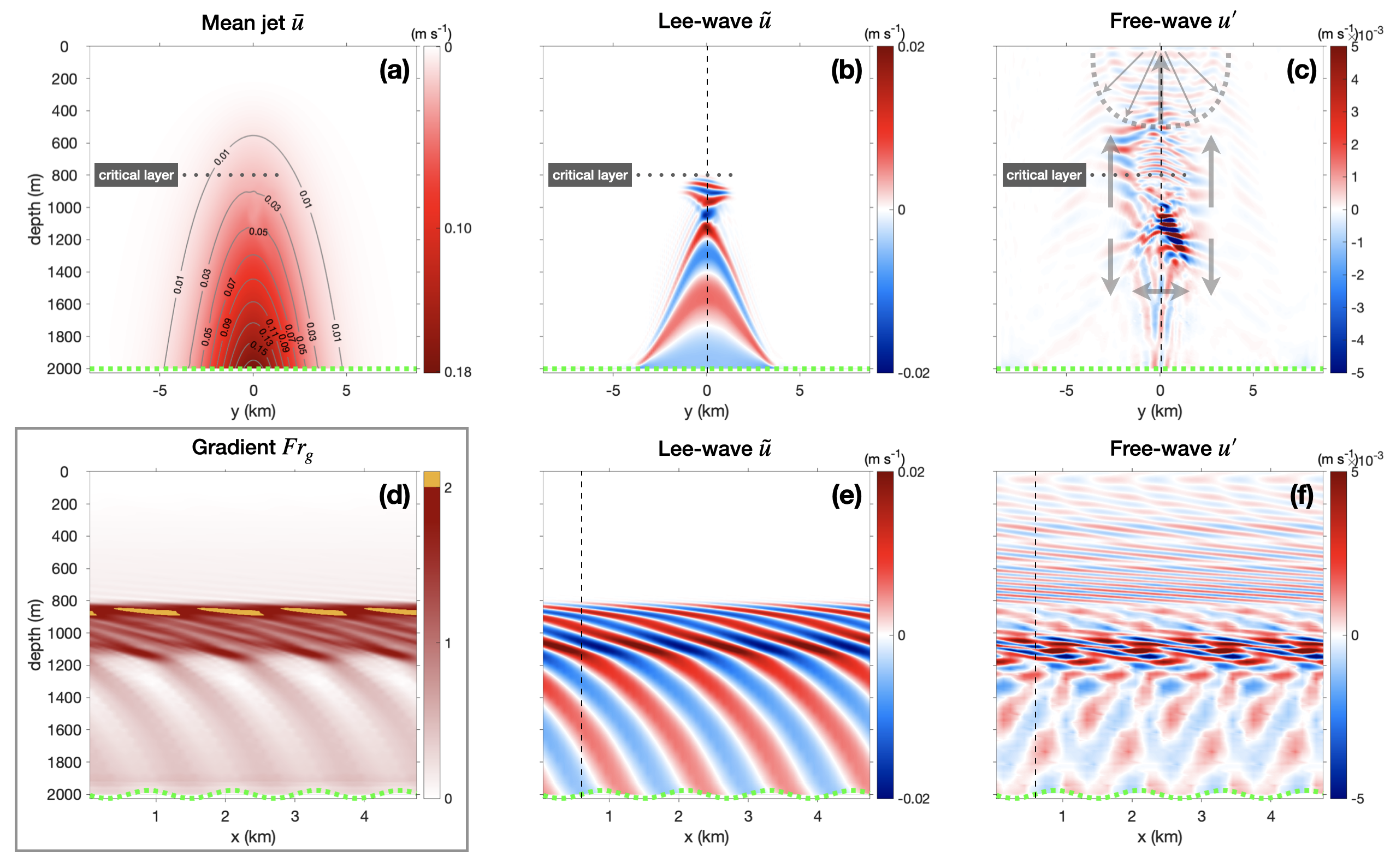}
 \caption{Meridional sections at $x=0.6$ km (upper row) and zonal sections along the jet axis at $y=0$ (lower row) of zonal velocities for (a) the mean, (b,e) lee waves, and (c,f) free waves. Gradient Froude number $Fr_g=\sqrt{(\partial u/\partial z)^2+(\partial v/\partial z)^2}/N$ (d), where $Fr_g>2$ at 900-m depth (yellow) indicates potential shear instabilities. 
 In (c), grey arrows mark the direction of free-wave vertical and meridional group velocities, and grey dashed curve represents a wave front of surface-reflected IGWs. 
 The asymmetry about $y=0$ in (b) and (c) arises because of different stratification and absolute vorticity on either side of the jet.
 Grey dotted vertical lines in (b), (c), (e) and (f) mark where the zonal and meridional sections are taken. 
 Bottom topography is shown in green dotted curves amplified by a factor of 5 to be visible.} \label{fig_u}
\end{figure*}

In the first five days, we observe the development of lee waves. From days 5-15, a quasi-steady state is achieved that is used for the analysis. After day 15, the simulation becomes unsteady as the jet and lee waves slowly distort due to nonlinearity so that restoration cannot maintain the jet. Linear lee waves with dominant intrinsic frequency $|k_0U_0|\approx 0.94N$ are generated by bottom topography of wavenumber $k_0$ and propagate upward in the bottom-intensified jet, becoming increasingly nonlinear with height above bottom as vertical wavenumber $m = k \sqrt{(N^2-k^2U^2)/(k^2U^2-f^2)}$ increases with decreasing $U$ (Figure~\ref{fig_u}b,e). 
A critical layer is expected at 800-m depth where $|k_0U| \approx |f|$ and $m\rightarrow \infty$. 
Strong vertical shear induced by nonlinear lee waves has the potential to overcome density stratification and trigger shear instabilities where gradient Froude number $Fr_g=\sqrt{(\partial u/\partial z)^2+(\partial v/\partial z)^2}/N$ exceeds 2 (Figure~\ref{fig_u}d; \citealt{Miles1961}). 

Internal gravity waves (IGWs) with temporally varying phases (free waves) are observed (Figure~\ref{fig_u}c,f). 
Unlike the stationary lee waves trapped within the jet (Figure~\ref{fig_u}b), free waves can potentially radiate out of the jet to reflect downward from the free surface (Figure~\ref{fig_u}c) or inward from the meridional boundaries.
In the meridional section (Figure~\ref{fig_u}c), temporally-varying phases of free waves propagate toward 1100-m depth from above and below, signifying that energy radiates outward from this depth. 
Below 1200-m depth, free-wave phases propagate in both meridional directions within the jet (Figure~\ref{fig_u}f). 
Free waves do not reach the lateral boundaries because the biharmonic damping dissipates them.

\begin{figure*}[hbp]
\centering
 \noindent\includegraphics[width=.5\textwidth,angle=0]{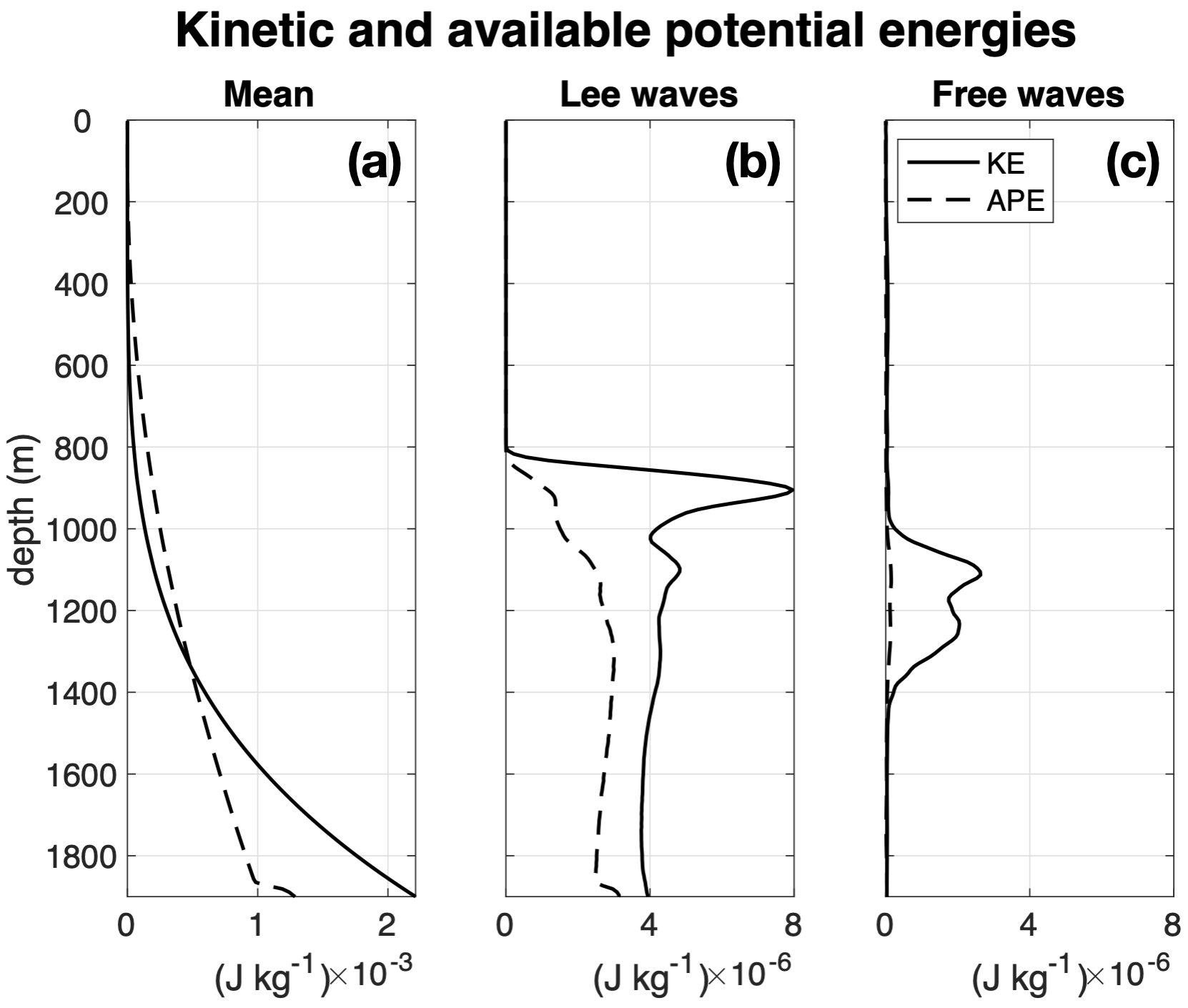}
 \caption{Kinetic (solid) and available potential energies (dashed) for (a) the jet, (b) lee waves and (c) free waves. Kinetic energies are dominated by horizontal speeds $|u+iv|$ for all three fields. Note the different energy scale in (a).} \label{fig_energy}
\end{figure*}

Free-wave kinetic and available potential energies both peak at roughly 1100-m depth (Figure~\ref{fig_energy}c). They have high vertical wavenumbers. Similar oscillations were reported in numerical simulations by \cite{Nikurashin2014} and \cite{Zemskova2021} for topographic Froude numbers exceeding 0.7 for 1-D and 0.4 for 2-D topography. 
Their high kinetic-to-available-potential-energy ratio indicates that they are near-inertial waves. 
The escaped fraction of wave energy above 800-m depth is only 1\%, making remote dissipation unable to explain the $O(1)$ dissipation deficit \citep{Waterman2014}. As a consequence, hypothesis (ii) that generation of free waves can lead to remote dissipation can be ruled out.

\subsection{Spectral analysis}
\begin{figure*}[hbp]
\centering
 \noindent\includegraphics[width=.8\textwidth,angle=0]{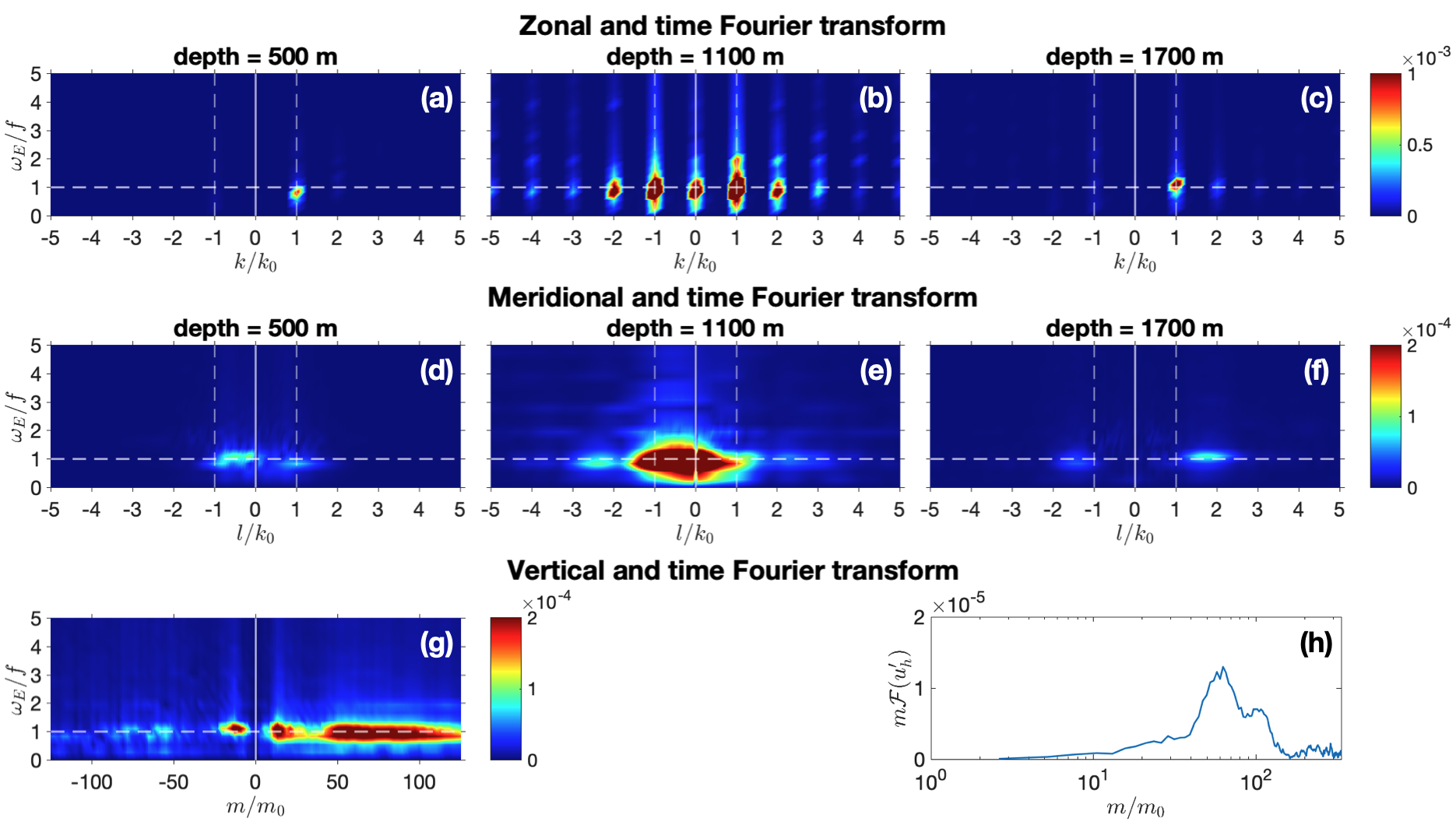}
 \caption{(a--g) Two-dimensional signed Fourier transforms of free-wave horizontal velocity $u'_h=u'+iv'$ (with mean and lee-wave components removed) across the middle of the jet but at different locations. 
 Eulerian frequency $\omega_E$ is normalized by the absolute value of the Coriolis frequency $|f|=1.3\times 10^{-4}$ rad s$^{-1}$, zonal and meridional wavenumbers $(k,l)$ by topographic wavenumber $k_0=5.23\times 10^{-3}$ rad m$^{-1}$, and vertical wavenumbers $m$ by the bottom lee-wave vertical wavenumber $m_0=k_0\sqrt{(N^2-k_0^2U_0^2)/(k_0^2U_0^2-f^2)}=1.88\times 10^{-3}$ rad m$^{-1}$.
 White solid lines mark $k=l=m=0$, and white dashed lines mark $\omega_E/f=1$, $k/k_0=\pm 1$, and $l/k_0=\pm 1$. 
 (h) Variance-preserving 1-D vertical wavenumber slice of (g) along $\omega_E=f$. 
 } \label{fig_fft}
\end{figure*}

Zonal and temporal Fourier transforms of free-wave horizontal velocity $\mathbf{u}_h'=u'+iv'$ show that free waves are near-inertial in the Eulerian frame and peak at the topographic wavenumber $k_0$ (Figure~\ref{fig_fft}a,b,c). 
At $z=1100$ m where free-wave energy peaks, the spectra stand out at $k=0$ and superharmonics of $\pm k_0$ (Figure~\ref{fig_fft}b). 
The $k=0$ and super-harmonic near-inertial oscillations have high vertical wavenumber and zero group velocities so are not found shallower or deeper in the water column.
At this depth, lee waves ($\omega_E^{\text{lee}}=0$) have intrinsic frequency $\omega_I^{\text{lee}}=|\omega_E^{\text{lee}}-k_0 U(z)|=2f$. The $k=0$ and $k=k_0$ free waves have both Eulerian and intrinsic frequencies at inertial frequency because of Doppler-shifting: $\omega_I^{\text{free}}=|\omega_E^{\text{free}}-0 \cdot U(z)|=f$ and $\omega_I^{\text{free}}=|\omega_E^{\text{free}}-k_0U(z)|=|f-2f|=f$.
Lee waves and free waves can have resonant triad interactions through parametric subharmonic instability (PSI). 
PSI is the decay of a low-vertical-wavenumber parent wave into two nearly identical high-vertical-wavenumber daughter waves with half the parent-wave intrinsic frequency. As a frequency-halving mechanism, PSI is most effective in transferring energy at $2f$ toward $f$ so can contribute to the inertial peak of the internal wave field \citep{McComas1977a,Carter2006,MacKinnon2005}.
In our simulation, the dominant nonlinear wave-wave interaction is PSI, where lee waves serve as the parent wave and inertial free waves as the daughter waves. Such nonlinear interaction moves energy out of lee waves into high vertical wavenumbers and inertial frequencies. The rate of transfer is encapsulated in the lee-wave self-advection term with $-\frac{\partial}{\partial z} \left(\frac{1}{2} \overline{ \tilde{u} \tilde{u} \tilde{w}}\right) \sim O(10^{-10}$ W kg$^{-1}$) (Fig. 4b3) being the dominant component. 
This same mechanism may be responsible for the inertial oscillations that appeared in the lee-wave generation simulations of \cite{Nikurashin2014} and \cite{Zemskova2021} since their broadband topography will have included generation of $|kU_0| = 2f$ lee waves susceptible to PSI.

Meridional and temporal Fourier transforms also display near-inertial peaks (Figure~\ref{fig_fft}d,e,f). 
At $z=500$ and 1100 m (Figure~\ref{fig_fft}d,e), the spectrum spreads over $0 \leq |l/k_0| \lessapprox 1$, consistent with the lower meridional wavenumbers above 1200-m depth (Figure~\ref{fig_u}c). 
The spectra are broadband for $l>0$ and $l<0$ indicating two groups of IGWs with opposite meridional phase velocities (Figure~\ref{fig_u}c). 
At $z=1700$ m (Figure~\ref{fig_fft}f), the spectrum is dominated by higher meridional wavenumbers $1 \lessapprox |l/k_0| \lessapprox 2$, consistent with the steeper phase lines near the bottom (Figure~\ref{fig_u}c). 

Vertical and temporal Fourier transforms (Figure~\ref{fig_fft}g) show that free waves with $m>0$ dominate, indicating stronger downward than upward energy propagation. Most free waves are reflected downward either by the critical layer or the free surface (Figure~\ref{fig_u}c). Their propagation toward stronger mean flow allows extraction of jet energy by the dominant exchange term $-\frac{\partial U}{\partial z} \overline{\langle u' w'\rangle}\sim O(10^{-11}$ W kg$^{-1}$) through wave action conservation (Fig. 4c1). A variance-preserving vertical wavenumber slice of the Fourier transform of the horizontal free-wave velocity along $\omega_E=f$ (Figure~\ref{fig_fft}h) shows that most free waves are at high vertical wavenumber $m \gg m_0$ where $m_0$ is the lee-wave vertical wavenumber at the bottom, consistent with the PSI mechanism which is an ultraviolet catastrophe that moves energy to the smallest scale possible until viscosity takes over. 


\subsection{Energy budgets}
\begin{figure*}[bp]
\centering
 \noindent\includegraphics[width=\textwidth,angle=0]{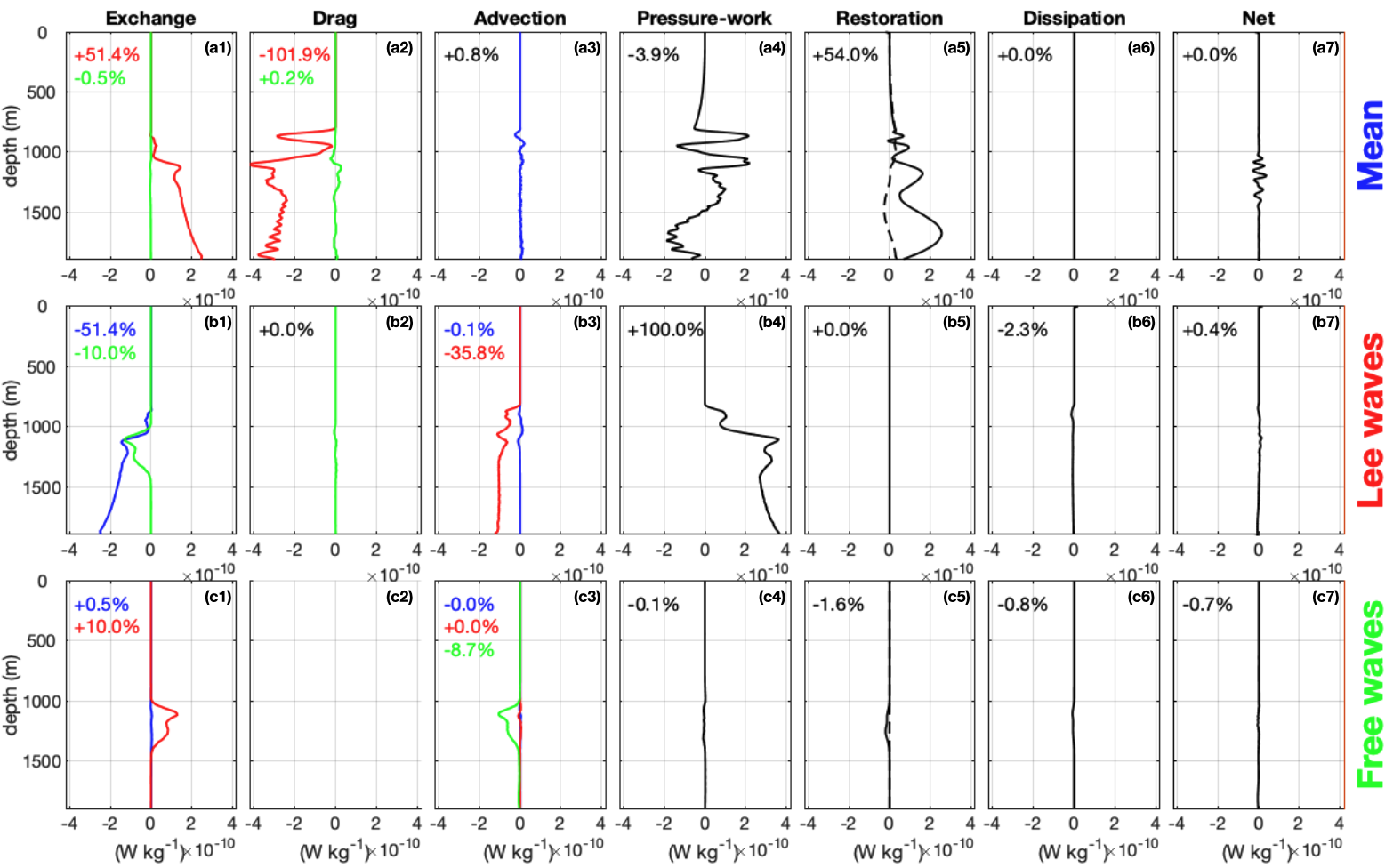}
 \caption{Energy budgets for (a1--7) the mean, (b1--7) lee waves and (c1--7) free waves based on energy conservation equations \eqref{M3}, \eqref{L3}, and \eqref{F3}. 
 In (a5) and (c5), solid and dashed curves show the restoration of kinetic and available potential energies, respectively. Percentages on each panel show the integrated energy input (positive) or output (negative) for each field and is normalized by lee-wave generation (lee-wave pressure-work; b4). Note that the horizontal axes are zoomed in by a factor of 10 for free waves compared to those for the mean and lee waves. Blue, red, and green mark components associated with the mean, lee waves, and free waves, respectively.} \label{fig_budget}
\end{figure*}

\begin{figure*}[bp]
\centering
 \noindent\includegraphics[width=.9\textwidth,angle=0]{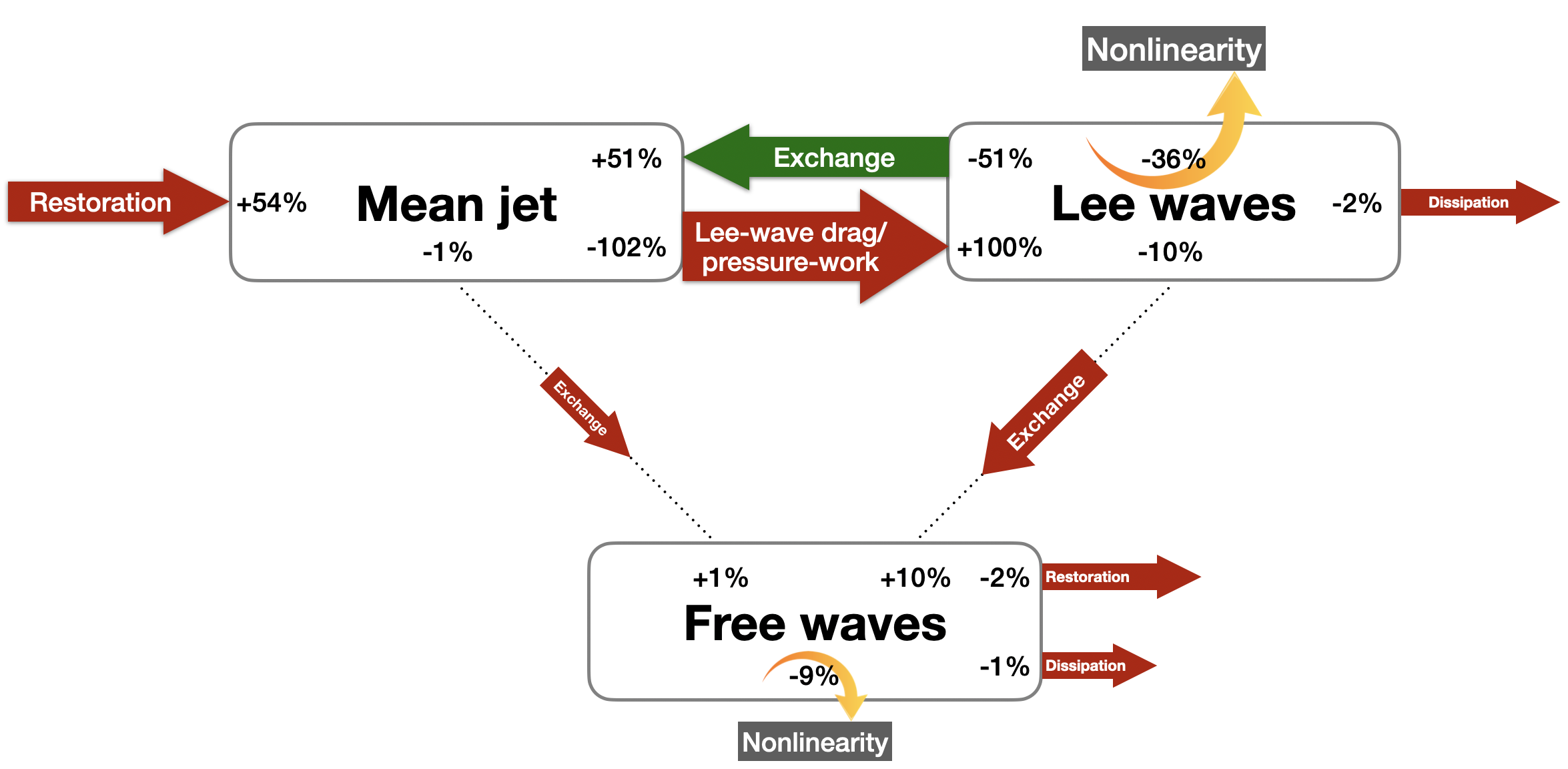}
 \caption{Energy budgets for (a1--7) the mean, (b1--7) lee waves and (c1--7) free waves based on energy conservation equations \eqref{M3}, \eqref{L3}, and \eqref{F3}. Percentages on each panel show the integrated energy input (positive) or output (negative) for each field and is normalized by lee-wave generation (lee-wave pressure-work; b5).} \label{fig_triangle2}
\end{figure*}

At quasi-steady state, the kinetic and available potential energies in the mean, lee-wave and free-wave fields do not vary in time so energy sources and sinks through the six forcings in \eqref{M3}, \eqref{L3} and \eqref{F3} balance. 
Decomposed energy budgets are averaged in the horizontal (Figure~\ref{fig_budget}). 
The leading-order mean budget is fueled by exchange with lee waves (i.e., reabsorption; 51\%) and restoration (54\%), and drained by lee-wave drag at the bottom (-102\%); all percentages are normalized by lee-wave generation [also known as lee-wave pressure-work; Figure~\ref{fig_budget}(b4)] for comparison.
The lee-wave budget gains from pressure-work due to bottom generation (100\%), and loses to exchange with the mean (-51\%), exchange with free waves (-10\%), nonlinear transfer (i.e., lee-wave self-advection; -36\%) and dissipation (-2\%). 
The free-wave budget gains through exchange with lee waves (10\%) and with the mean (1\%), and loses to nonlinear transfer (i.e., free-wave self-advection; -9\%), dissipation (-1\%) and restoration of $k=0$ free waves (-2\%).
The net budgets for all three fields are within $\pm 1\%$ of lee-wave generation, indicating closure [Figure~\ref{fig_budget}(a7), (b7) and (c7)]. 

Energy budgets are summarized in Figure~\ref{fig_budget}. Bottom generation converts energy from the mean jet into lee waves [Figure~\ref{fig_budget}(a2; b4)]. As expected, lee-wave drag on the mean flow almost balances lee-wave pressure-work at the bottom, i.e., $\overline{\tilde{w}\tilde{p}} = -\left[\overline{\tilde{u}\tilde{w}} - (f/N^2)\overline{\tilde{v}\tilde{b}}\right] U_0 \approx -\overline{\tilde{u}\tilde{w}} U_0$, where $\overline{\tilde{u}\tilde{w}} - (f/N^2)\overline{\tilde{v}\tilde{b}}$ is the \cite{Eliassen1960} flux and the cross-stream buoyancy-flux is negligible.
During upward lee-wave propagation in the water column, $\sim 50\%$ of the bottom generation is reabsorbed back to the mean [red curve in Figure~\ref{fig_budget}(a1)] via the dominant exchange term $\frac{\partial U}{\partial z} \overline{\tilde{u}\tilde{w}}$ due to wave action conservation \citep{Kunze2019,Wu2023}, while restoration injects another $\sim 50\%$ into the mean to maintain the jet [Figure~\ref{fig_budget}(a5)]. 

For lee waves, in addition to losing $\sim 50\%$ of their generated energy back to the mean [reabsorption, blue curve in Figure~\ref{fig_budget}(b1)], another $\sim 36\%$ is lost through nonlinear wave-wave interactions, predominantly PSI encapsulated in the dominant lee-wave self-advection term $-\frac{\partial}{\partial z} \left(\frac{1}{2}\overline{\tilde{u}\tilde{u}\tilde{w}}\right)$ [red curve in Figure~\ref{fig_budget}(b3)], $\sim 10\%$ to free waves through free-wave/lee-wave exchange between 1000--1400-m depth [green curve in Figure~\ref{fig_budget}(b1)], and only $\sim 2\%$ to explicit dissipation [Figure~\ref{fig_budget}(b6)].

Free waves receive $\sim 1\%$ and $\sim 10\%$ of generated lee-wave energy from the mean and lee waves via the dominant exchange terms $-\frac{\partial U}{\partial z} \overline{\langle u'w' \rangle}\sim O(10^{-11}$ W kg$^{-1})$ and $-\overline{\frac{\partial \tilde{u}}{\partial z} \langle u'w' \rangle} - \overline{\frac{\partial \tilde{v}}{\partial z} \langle v'w' \rangle}\sim O(10^{-10}$ W kg$^{-1})$, respectively [Figure~\ref{fig_budget}(c1)]. 
The $\sim 1\%$ energy gain from the mean is due to wave action conservation since the free waves propagate predominantly downward toward greater $U$ (Figure~\ref{fig_u}c). 
In contrast to the mean jet whose energy is replenished through restoration [54\%; Figure~\ref{fig_budget}(a5)], free waves lose energy to restoration [-2\%; Figure~\ref{fig_budget}(c5)], because the zonal average of the total zonal velocity and density fields are restored to their initial conditions, suppressing growth of the $k=0$ free waves. 
Explicit dissipation of free waves is small [-1\%; Figure~\ref{fig_budget}(c6)], while most free-wave energy is drained through nonlinear transfer [-9\%; Figure~\ref{fig_budget}(c3)] via the dominant free-wave self-advection term $-\frac{\partial}{\partial z} \left(\frac{1}{2}\overline{u'u'w'}+\frac{1}{2}\overline{v'v'w'}\right) \sim O(10^{-10}$ W kg$^{-1})$. 

\subsection{Nonlinear transfer as indirect energy loss}

Nonlinear transfer represented by lee-wave and free-wave self-advection terms (i.e., the two higher-order terms in the wave energy equations) are significant energy sinks in the budget. This interpretation may seem counter-intuitive since advection is conventionally viewed as a mechanism that shuffles energy in physical and/or spectral space. However, parametric subharmonic instability (PSI) is the dominant mechanism of nonlinear interactions here, transferring energy predominantly from $2f$ to $f$ and to high vertical wavenumber (Fig.~3). PSI is an ultraviolet nonlocal process involving vertically scale-separated parent and daughter waves so cannot be fully resolved by the numerical model. The transfer of energy to unresolved small scales by this process results in indirect dissipation. 
The 45\% nonlinear sink in this study consists of contributions from the lee-wave self-advection term in the water column [36\%; $-\frac{\partial}{\partial z} \left(\frac{1}{2}\overline{\tilde{u}\tilde{u}\tilde{w}}\right)$; red curve in Figure~\ref{fig_budget}(b3)] and the free-wave self-advection term near the critical layer [9\%; $-\frac{\partial}{\partial z} \left(\frac{1}{2}\overline{u'u'w'}+\frac{1}{2}\overline{v'v'w'}\right)$; green curve in Figure~\ref{fig_budget}(c3)]. All percentages are normalized by lee-wave generation at the bottom. 

From the perspective of numerical modeling, finite-difference treatment (as well as the finite-volume method employed in PSOM) of the advection terms introduces discretization errors that act as dissipation and diffusion that implicitly remove energy from the system even in the absence of explicit dissipation. These discretization errors are most pronounced at small scales and for high nonlinearity. This indirect dissipation explains why the nonlinear sink increases from 30\% in the more viscous and less nonlinear case of \cite{Wu2023} to 45\% with reduced viscosity and higher nonlinearity in this paper (Figure~\ref{fig_pathway}). 

\section{Conclusions and discussion}
 
\begin{figure*}[bp]
\centering
 \noindent\includegraphics[width=\textwidth,angle=0]{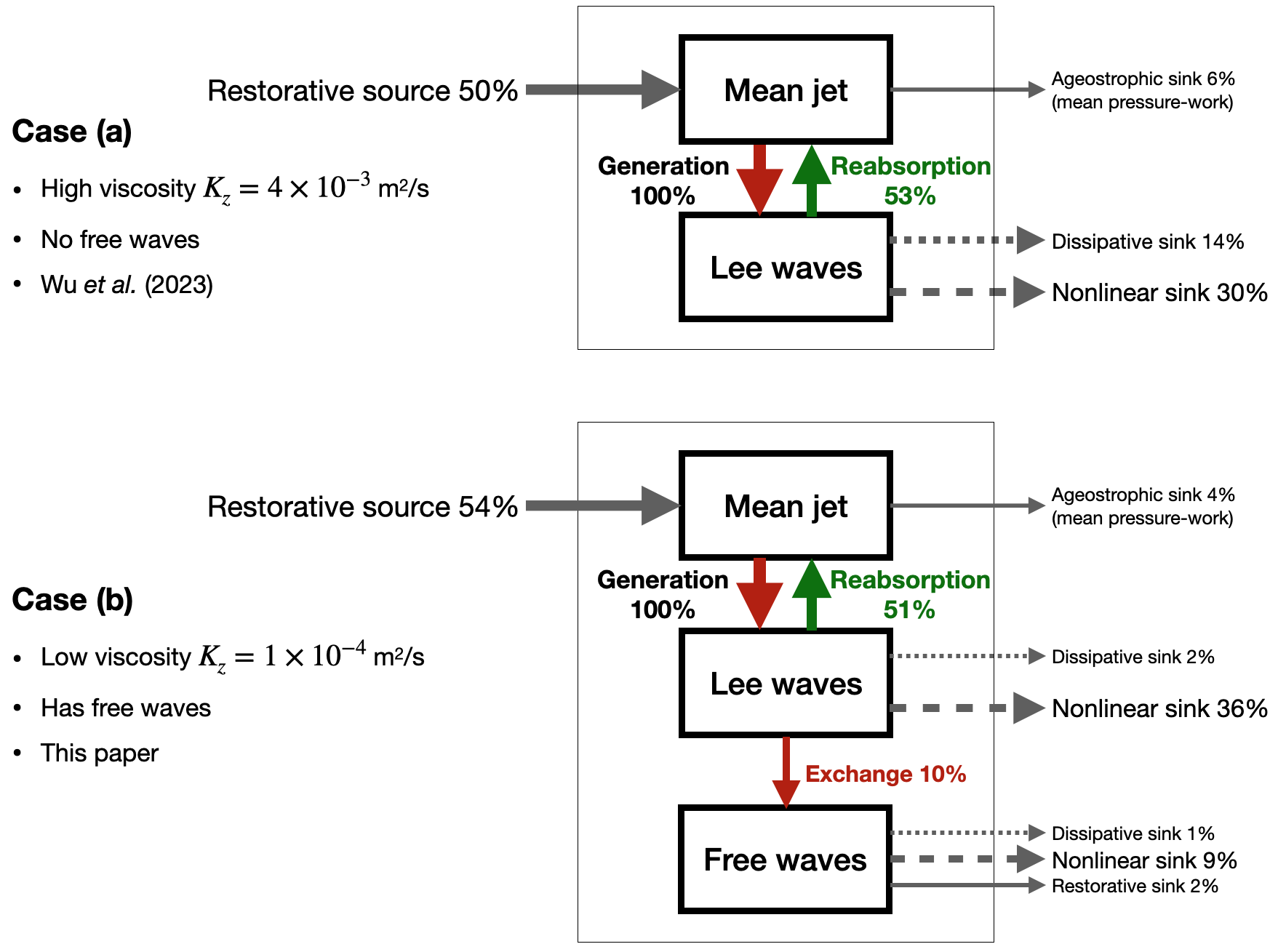}
 \caption{Energy pathways for cases (a) without free waves due to strong damping \citep{Wu2023} and (b) with free waves (this paper). Arrows indicate the direction of energy transfer with arrow size approximately proportional to the rate of transfer. Percentages show the rate of transfer normalized by lee-wave generation (lee-wave pressure-work). In between the dominant fields (i.e., jet, lee- and free waves), red arrows indicate forward cascade and green backward. } \label{fig_pathway}
\end{figure*}

\begin{figure*}[bp]
\centering
 \noindent\includegraphics[width=.6\textwidth,angle=0]{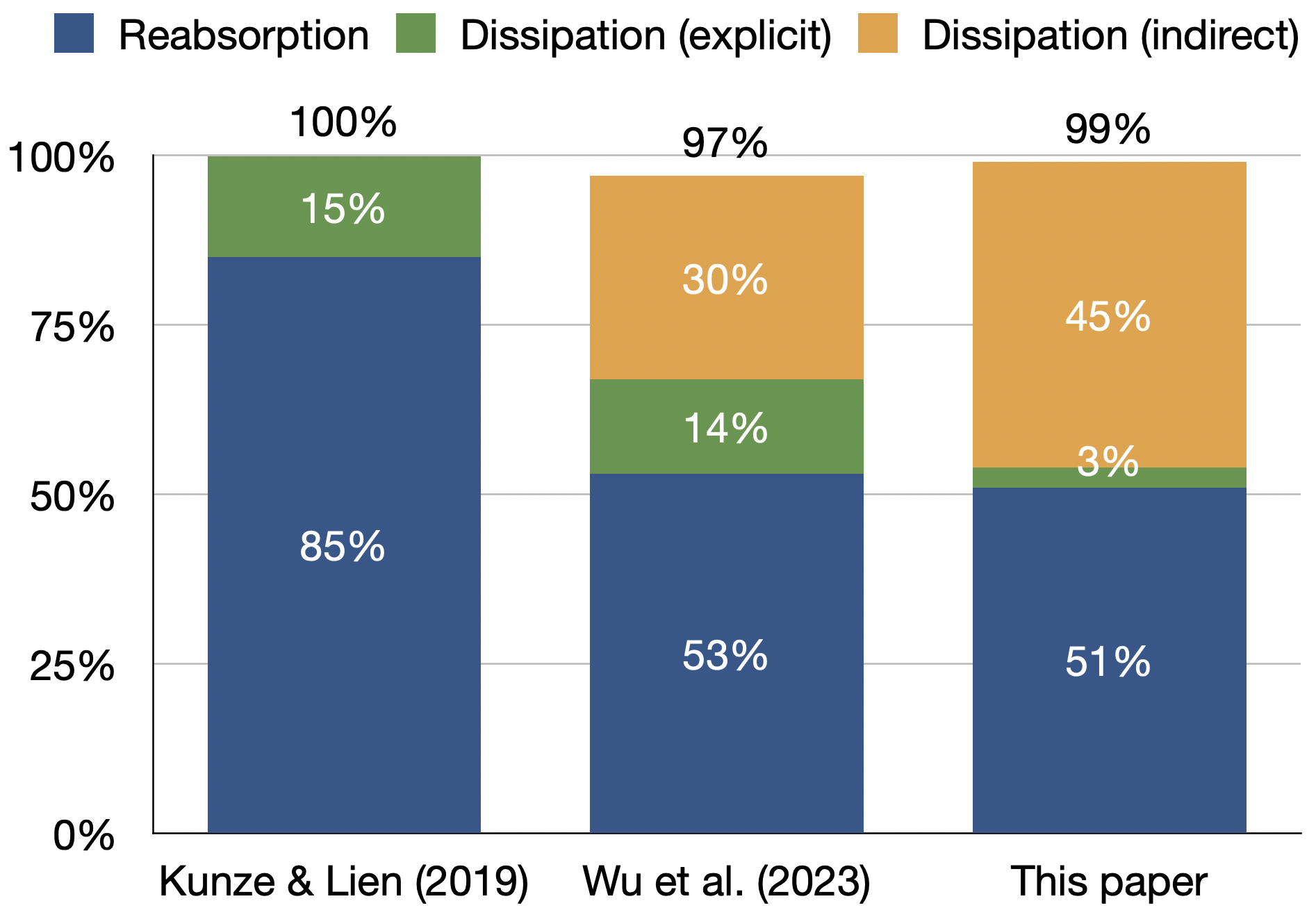}
 \caption{Partition of lee-wave energy sinks for (a) wave action conservation prediction by \cite{Kunze2019}, (b) numerical simulations using strong subgridscale damping \citep{Wu2023} and (c) using weak damping (this paper). Reabsorption is blue, explicit dissipation green, indirect dissipation due to nonlinear transfer orange.} \label{fig_bar}
\end{figure*}

Freely-propagating internal gravity waves (free waves) are emitted from local shear instability at a single critical layer of steady monochromatic lee waves in bottom-intensified flow. 
Triple-decomposition of energy equations into the zonal mean, lee-wave and free-wave components isolates both wave-mean and wave-wave interactions and can be used as a general tool of budget analysis with broader applications to eddy-wave and eddy-mean interactions.

Free waves excited by shear instability at a critical layer could, in principle, lead to remote dissipation \citep{Kunze1995, Wright2014}. These free waves can radiate out of the jet to leave a signature in the upper water column. But the escaped fraction is only 1\% so insufficient to explain the observed turbulence shortfall in microstructure measurements at two sites of strong lee-wave generation in the Southern Ocean (e.g., \citealt{ Waterman2014}), indicating that radiation of free waves is not a significant sink, at least for the case considered here.

Energy pathways for cases without free waves \citep{Wu2023} and with free waves (this paper) are compared in Figure~\ref{fig_pathway}. The reabsorbed fraction of lee-wave energy in bottom-intensified flow is $\sim 50\%$ in both cases. The explicit dissipative fraction shrinks from 14\% to 3\% (2\% for lee waves and 1\% for free waves) as a direct consequence of reduced eddy viscosity and diffusivity, while the nonlinear fraction increase from 30\% to 45\% (Figure~\ref{fig_bar}). 
Interpreting the nonlinear sinks as ultraviolet indirect dissipation, the total dissipation fraction (explicit + indirect via nonlinear transfer) is $\sim 50\%$ in both cases. Thus, the partition of lee-wave sinks into reabsorption and dissipation is independent of the turbulent parameterization employed in numerical models. The remaining degrees of freedom to be considered in future research are the topographic wavenumber and steepness, that is, the horizontal wavenumber spectrum of topographic height $h$, since these parameters determine lee-wave intrinsic frequency and radiation at generation. This parameter space is being explored by the authors.

The nonlinear sinks are shown to be significant in the above cases of linear lee-wave generation by gentle topography. In cases of nonlinear generation by steeper topography, the nonlinear sinks are anticipated to be potentially more prominent.
In the ocean, nonlinear transfer acts as a major mechanism for turbulence production and dissipation of internal gravity waves (e.g., \citealt{McComas1981,Henyey1986,Gregg1989,Eden2019a,Eden2019b,Eden2020,Pan2020,Dematteis2021,Dematteis2022,Wu2023b}). This is absent from linear wave action conservation \citep{Kunze2019} and was not correctly interpreted in the steady lee-wave numerical simulations of \cite{Wu2023}. That is, the predicted dissipative fraction $|f/kU_0|$ from wave action conservation underestimates true lee-wave dissipation. Nonlinear transfer to dissipation represents an additional turbulent loss term, reducing the reabsorptive fraction and increasing the dissipative fraction relative to wave action conservation predictions, thus increasing the role of lee waves in dissipating the balanced circulation. However, the results found here are for idealized monochromatic topography only. The authors will explore the partition between lee-wave reabsorption and dissipation for more realistic broadband topography as a next step towards parameterizing the role of lee waves in dissipating balanced circulation.

\acknowledgments
This research was supported by NSF Grant OCE-2306124 (UMich), OCE-1756279 (WHOI), OCE-1756093 (NWRA), and OCE-1755313 and OCE-2148404 (UMass Dartmouth). 

\datastatement

The Process Study Ocean Model is illustrated on the lab webpage at https://mahadevan.whoi.edu/PSOM. Model configuration and code has been uploaded as an example experiment in the GitHub archive of the PSOM Version 1.0 at https://github.com/PSOM/V1.0/tree/master/code/leewaves. Simulation outputs and analysis on which this paper is based are too large to be retained or publicly archived with available resources but will be made available to collaborators or interested individuals upon request.

%







\bibliographystyle{ametsocV6}
\bibliography{library,library_added}

\end{document}